# Gravitational anomaly in ferrimagnetic topological Weyl semimetal NdAlSi

Pardeep Kumar Tanwar [1], Mujeeb Ahmad [1], Md Shahin Alam [1], Xiaohan Yao [2], Fazel Tafti [2] and Marcin Matusiak [1,3*]

1. *International Research Centre MagTop, Institute of Physics, Polish Academy of Sciences, Aleja Lotnikow 32/46, PL-02668 Warsaw, Poland*
2. *Department of Physics, Boston College, Chestnut Hill, Massachusetts, 02467, USA*
3. *Institute of Low Temperature and Structure Research, Polish Academy of Sciences, ul. Okólna 2, 50-422 Wrocław, Poland*

Quantum anomalies are the breakdowns of classical conservation laws that occur in quantum–field theory description of a physical system. They appear in relativistic field theories of chiral fermions and are expected to lead to anomalous transport properties in Weyl semimetals. This includes a chiral anomaly, which is a violation of the chiral current conservation that takes place when a Weyl semimetal is subjected to parallel electric and magnetic fields. A charge pumping between Weyl points of opposite chirality causes the chiral magnetic effect that has been extensively studied with electrical transport. On the other hand, if the thermal gradient, instead of the electrical field, is applied along the magnetic field, then as a consequence of the gravitational (also called the thermal chiral) anomaly an energy pumping occurs within a pair of Weyl cones. As a result, this is expected to generate anomalous heat current contributing to the thermal conductivity. We report an increase of both the magneto-electric and magneto-thermal conductivities in semi-classical regime of the magnetic Weyl semimetal NdAlSi. Our work also shows that the anomalous electric and heat currents, which occur due to the chiral magnetic effect and gravitational anomalies respectively, are still linked by a 170 years old relation called the Wiedemann-Franz law.

Corresponding author, email: [*]matusiak@magtop.ifpan.edu.pl

**Introduction**

Topological materials are a class of compounds having non-trivial electronic band structures [1-3]. Their characteristic linear energy dispersion and spin momentum locking lead to the emergence of new phenomena with potential applications in the field of spintronics and ultra-fast electronics devices [4]. Anomalous properties of topological materials have been extensively studied to date, and a list of investigated phenomena includes the chiral magnetic effect (CME) [5,6], anomalous Hall [7-10] and anomalous thermal Hall effects [11-13], chiral zero sound [14-16], mixed axial-gravitational anomaly [17-22] and more [22,23]. Some theoretical predictions have been well evidenced experimentally, some would benefit from alternative confirmation. Among the latter is the occurrence of the chiral anomaly in Weyl semimetals, which is expected to generate additional electric current, when electric and magnetic fields are parallel to each other [24-29]. This should contribute to the total electrical conductivity, hence an observation of the negative longitudinal magnetoresistance (NLMR) was initially taken as the smoking-gun evidence of the chiral anomaly [30,31]. However, it was later realised that the phenomenon could be also attributed to other factors, such as the current jetting effect [32-34] or the geometric-dependent magnetoresistance [35]. On the other hand, a charge pumping between Weyl nodes of opposite chirality is not the only anomalous behaviour of Weyl semimetals. Namely, in presence of a thermal gradient an energy pumping between Weyl nodes should also affect the entropy current and contribute both to the thermal conductivity [36] and the thermoelectric power [20,29,37]. These phenomena are supposed to be more robust to experimental artefacts [20-22,36]. Interestingly, the resulting anomalous thermal effect is recognized to be a solid-state realisation of the gravitational anomaly known from high-energy physics [19,20,22,38].

In this work, we show evidence for the presence of the gravitational anomaly in the semi-classical regime of type II magnetic Weyl semimetal NdAlSi. Previously, an increase of the thermal conductivity with magnetic field was only observed in the extreme quantum limit of $Bi_{1-x}Sb_x$, which a particular case of a field-induced "ideal" Weyl semimetal with the Fermi level matching exactly to the position of the Weyl points [36]. We measured the thermal and electrical transport in the magnetic field parallel to the thermal gradient (or electric field), and we observed an increase in the thermal (or electrical) conductivity, indicating the emergence of the chiral thermal (or electric) current. A relation between the thermal and electrical response remains consistent with theoretical predictions. Our study shows that the gravitational anomaly can be detected in a semi-classical regime with the Fermi level not matching exactly the position of Weyl nodes.

**Results and discussion**

NdAlSi is a magnetic type II Weyl semimetal in which both inversion and time-reversal symmetries are broken [39-42]. Its ground state is ordered ferrimagnetically, which, with increasing temperature, transforms into an antiferromagnetic order and eventually into a high temperature paramagnetic phases [39,40]. In each state, the electronic structure of NdAlSi hosts Weyl fermions with 26, 28 and 20 pairs of Weyl nodes in the ferrimagnetic, antiferromagnetic and paramagnetic phases, respectively [39,41]. A sizable Dzyaloshinkii – Moriya interaction between local magnetic moments leads to their helical magnetic structure [39,40,42,43]. Since the incommensurate wavevector connects different non-trivial Fermi pockets, the magnetism in NdAlSi appears to be mediated by Weyl fermions [40].

Figure 1a presents the magnetic field dependences of the electrical resistivity for NdAlSi measured at various temperatures with $B$ parallel to both the electric current and the $a$-axis. At low temperatures there are visible anomalies in $\rho(B)$ (like at $B \approx 2$, 10 T at $T \approx 4.7$ K, and $B \approx 12$ T at $T = 1.8$, 2.8 K), which presumably in high magnetic field marks a suppression of the spin density, as it closes the gap in the nested parts of the Fermi surface. The measurements were performed in increasing and decreasing magnetic field and we did not notice hysteretic effects in this field region. Inset in Fig. 1a shows the zero-field temperature dependence of the resistivity with the electric current applied along the *a*-axis. From room temperature, $\rho(T)$ gradually decreases down to $T_{incom}$ = 7.2 K, where it exhibits a kink corresponding to the paramagnetic – incommensurate spin density wave transition [39]. The incommensurate ordering wavevector decreases with decreasing temperature and at $T_{com} \approx 3.2$ K NdAlSi undergoes a transition to the commensurate chiral ferrimagnetic phase marked by another kink in $\rho(T)$ [39]. The increase of $\rho$ below $T \approx 5.5$ K may be due to a superzone gap formation that can develop in a case when the periodicity of antiferromagnetic order is different from that of the lattice [44-46]. The transitions temperatures found here are consistent with previous reports [39,40].

For all temperatures $\rho(B)$ increases with *B* in the low field region, which can be due to the positive orbital longitudinal magnetoresistance ($\sigma_0$) and/or the weak antilocalization (WAL) effect [47,48]. The former can be related to Fermi surface anisotropy or the momentum – dependent scattering time [49,50]. The presence of the latter is one of the transport signatures for Weyl semimetals [47,51-54] which could be related to quantum interference of Weyl fermions [53]. The 3D WAL is expected to contribute to the total magneto-electrical conductivity ($\sigma_{total}$) as $-\sqrt{B}$ or $-B^2$, depending on the strength of the magnetic field [53]. The $\sigma_{total}(B)$ can then be quantified phenomenologically as [54]:

$$\sigma_{total}(B) = \sigma_{ch} - C_{WAL}\left(\sqrt{B}\frac{B^2}{B^2+B_C^2} + \gamma B^2 \frac{B_C^2}{B^2+B_C^2}\right) + \sigma_0, \tag{1}$$

where $\sigma_{ch}$ is the anomalous chiral contribution, $C_{WAL}$ and *γ* are parameters describing WAL, $B_C$ characterises a crossover from $-\sqrt{B}$ to $-B^2$ regime ($B_c \approx \frac{\hbar}{el_\varphi^2}$; $\hbar$, $l_\varphi$ are the reduced Planck constant and dephasing length, respectively). The positive $\sigma_{total}(B)$ present at high temperature is unlikely due to WAL, indicating that $\sigma_0$ in NdAlSi is also field dependent. The exact form of $\sigma_0(B)$ is unknown, which makes difficult to separate the low-field positive magnetoresistance for specific contributions. Alternatively, in the high field and for temperatures lower than $T \approx 35$ K we observe the negative longitudinal magnetoresistance (NLMR), i.e. $\rho(B)$ exhibits a negative slope. It is unlikely related to domain walls as an application of a magnetic field of just over 1 mT causes the average area of stable domains in sister Weyl semimetal CeAlSi to more than double [55] and NLMR appears in the magnetic field about four orders of magnitude higher. Moreover, the effect is also present in the paramagnetic phase.

A decrease in the resistivity only when a magnetic field is applied parallel to the electrical current and along the separation of Weyl points is referred to as the chiral magnetic effect. This macroscopic manifestation of the chiral anomaly has been reported in the number of topological semimetals and is a consequence of an imbalance in the number of Weyl fermions populating Weyl nodes of opposite chirality that leads to anomalous electric current [5,28,30,37,54,56]. The resulting additional contribution to the longitudinal electrical conductivity in semi-classical regime and for $\mu >> T$ is:

$$\sigma_{ch} = N_W \frac{e^2}{8\pi^2 \hbar} \frac{(eB)^2 v^3}{\mathcal{E}_F^2} \tau_{WP}, \quad (2)$$

where $N_W, e, v, \mathcal{E}_F, \tau_{WP}$ are: number of Weyl nodes pairs, elementary charge, Fermi velocity, Fermi energy, and inter-valley Weyl relaxation time, respectively [57]. The latter, $\tau_{WP}$, determines the time scale at which quasiparticles scatter between the Weyl points and change their chirality. The increase of $\sigma$ should be then proportional to $B^2$, which is indeed the observed behaviour (both below and above $T_{incom}$) along with a small oscillatory component in the data presented in Fig. 1b. The frequency of these small oscillations is about 53 T (for more details, see Fig. S3 [60]) close to the $\beta$ frequency (66 T), dominating the high field Shubnikov – de Haas effect for $B \parallel c$ [40]. Using the Onsager relation $A = \pi k_F^2 = 2\pi eF/\hbar$ [58] (where $A$ is the area of the Fermi surface extreme cross-section, $k_F$ is the Fermi wavevector, and $F$ is the oscillations frequency), effective mass resulting from the Lifshitz-Kosevich formula (see inset in Fig. S3): $m^* = 0.11\, m_0$ ($m_0$ is the free electron mass) and mobility $\mu \approx 0.1$ m$^2$/Vs [40] we estimated $\tau_{WP}$ (Eq. 2) and the transport relaxation time $\tau_{tr} = \frac{\mu m^*}{e}$ for $T = 10$ K. As expected for the chiral regime, the resulting $\tau_{WP} \approx 10^{-11}$ s turns out to be significantly longer that $\tau_{tr} \approx 10^{-14}$ s, which is a necessary condition for observing CME [25].

However, the chiral magnetic effect is not the only possible origin of NLMR [33-36]. This can be also caused by extrinsic effects likely occurring in materials with ultra-high mobile charge carriers [32-34], but the mobility in NdAlSi is rather moderate, which makes this material less prone to extrinsic effects [59]. Nevertheless, the ultimate confirmation of the intrinsic nature of NLMR will be an observation of a counterpart phenomenon in the heat transport, which is robust to the current jetting effect [19,20,22,38].

Figure 2 presents the temperature dependence of the thermal conductivity ($\kappa$) along with the electronic thermal conductivity estimated using Wiedemann – Franz (WF) law: $\kappa_{WF} = \sigma L_0 T$, where $L_0$ is the Sommerfeld value of Lorenz number for free electrons ($\kappa(T)$ dependences measured at $B = 0, 5, 10$ and $14.5$ T are presented in Fig. S1 in Supplemental Material [60]). A maximum of $\kappa(T)$ at $T \approx 60$ K is most likely related to the lattice thermal conductivity and marks the temperature, at which Umklapp processes start to effectively disturb the phonon transport [61]. Below $T \approx 60$ K both $\kappa(T)$ and $\kappa_{WF}(T)$ decrease down to $T \approx 10$ K and at this temperature the electronic contribution appears to account for a significant portion of the total thermal conductivity. Below $T \approx 10$ K $\kappa(T)$ starts to deviate upwards and reaches a maximum at $T \approx 5$ K, which indicate that in this region magnons also participate in the heat transport. Inset in Fig. 2 shows the $\kappa(T) - \kappa_{WF}(T)$ data plotted along with the estimated phonon contribution of thermal conductivity, $\kappa_l(T)$. The latter was estimated using the low-temperature constant-pressure specific heat ($C_p$) data of non-magnetic reference material LaAlSi, which were fitted with the function $C_p(T) = \gamma T + aT^3$, to separate the specific heat for the electronic ($\gamma T$) and lattice ($aT^3$) contributions. Subsequently, the electronic contribution was subtracted from $C_p(T)$ and the remaining lattice specific heat, $C_l(T)$, was scaled to match the $\kappa(T) - \kappa_{WF}(T)$ of NdAlSi in the 18 – 23 K temperature range. If the phonon mean-free path (*l*) is constant below $T \approx 23$ K, then the resulting curve should roughly account for $\kappa_l(T)$, because $\kappa_l = 1/3\, C_v\, v_s\, l$ and we can assume that in this temperature range the speed of sound ($v_s$) does not change significantly and the constant-volume specific heat $C_v \approx C_p$. A low temperature extrapolation of the $\kappa_l(T)$ dependence thus obtained, shows that in this region $\kappa_l$ is much smaller than total thermal conductivity, while the maximum in $\kappa(T)$ at $T \approx 5$ K comes from the magnonic contribution to the thermal conductivity ($\kappa_m$) of NdAlSi mentioned earlier.

An important question to address is to what extent the Wiedeman-Franz law can be used to calculate the electronic thermal conductivity of NdAlSi based on its electrical conductivity. In general, the WF law is valid for a Fermi liquid, as long as collisions of charge carriers can be described as elastic, which means that the heat and charge currents are affected equivalently [60]. Hence, the WF law is usually well obeyed in the high and low temperature limits. Figure 3 presents the magnetic field dependences of $\kappa$ and $\kappa_{WF}$ for two exemplary temperatures within the paramagnetic phase of NdAlSi. Apparently, at room temperature $\kappa_{WF}(B)$ well accounts for the field variation of the $\kappa(B)$, if a contribution from field-independent phonon thermal conductivity, $\kappa_l(300\text{ K}) = 12.65$ W m$^{-1}$ K$^{-1}$, is taken into account. However, the same cannot be said for $\kappa_{WF}(20\text{ K})$, which increases in the high field limit, whereas $\kappa(B)$ at this temperature decreases monotonically in the entire field range. A likely explanation for this discrepancy lies in no longer field-independent phonon contribution. The phonons transport is often assumed to be not affected by a magnetic field, but this is not always the case. Interestingly, there are even examples of the phonon-based thermal Hall effect occurring in non-conductive materials [62,63]. In NdAlSi, the field-dependent phonon thermal conductivity originates likely from field dependent phonon scattering efficiency and we point at two possible underlying origins: an anharmonic phonon-phonon scattering [64] and phonon scattering on the paramagnetic free spins [65]. In the former, phonon induces the diamagnetic moment on atoms, which affects the orbital motion of valence band electrons. This phonon-induced diamagnetic moment alters the interatomic forces and leads to the magnetic-field-sensitive bond anharmonicities affecting the phonon-phonon interactions [64]. In the latter, the field sensitivity of phonon scattering is related to in-field splitting of the ground state of paramagnetic free moment $Nd^{+3}$. The lifting of its degeneracy leads to the two-level Schottky anomaly detected in the specific heat data [39]. The scattering of a phonon takes place in such a way that firstly, a phonon with energy equal to the energy difference between the split levels is absorbed, which excites free spin from the lower-energy state to the higher-energy. Later, another phonon is emitted by the excited state of free spin, which has the same energy but with a different wave vector.

Figure 4 presents a comparison of the magnetic field dependences of the total thermal conductivity $\kappa(B)$ to the respective $\kappa_{WF}(B)$ for several different temperatures in the low–$T$ region. The characteristic feature of $\kappa(B)$ is its initial steep decrease, which we attribute to the suppression of the magnonic component that is expected in the noninteracting approach to decrease approximately exponentially in the magnetic field [66,67]. The dependences of the thermal conductivity on the magnetic field after subtracting $\kappa_{WF}(B)$ do indeed exhibit such a behaviour and $\kappa_m(B)$ obtained in this way can be well fitted with the exponential decay function $e^{-cB}$ (see Fig. S2 [60]). Inset in Fig. 4 shows that for $T = 5$ and 6.1 K, a sum of $\kappa_{WF}(B)$ and the exponentially decaying magnonic conductivity matches very well the experimental $\kappa(B)$ data. This also indicates that magnons do not participate in thermal transport in high magnetic field limit [67,68].

Remarkably, below $T \sim 8$ K and in the high field regime, where the phonon and magnon contributions to the thermal conductivities are negligible, both $\kappa(B)$ and $\kappa_{WF}(B)$ consistently increase in a very good agreement with the Wiedemann-Franz law. At temperatures $T = 8$ K and 10 K, $\kappa_{WF}(B)$ eventually rises in the high field limit over $\kappa(B)$, which is likely due to a downward deviation of the Lorenz number below the Sommerfeld value. This is an expected consequence of the increase in temperature and the resulting difference in effectiveness of dissipation processes affecting heat and charge currents observed previously in metals and also topological semimetals [69,70]. An increasing role of inelastic scattering at high temperatures can also cause the positive thermal conductivity to be more difficult to observe. However, at low temperatures, where we expect a recovery of dominance of the elastic scattering, the high-field agreement between $\kappa(B)$ and $\kappa_{WF}(B)$ is almost perfect.

A Weyl system, which is subjected to the parallel magnetic and electric field, is expected to generate the additional electric current due to imbalance in number of Weyl fermions of opposite chirality [2,3,5,21]. On the other hand, if the thermal gradient is applied instead of the electrical field to such a material, there appears an imbalance in energy between two Weyl points of opposite chirality. The energy pumping that occurs between them leads to generation of the anomalous heat current [20,21,22,36], which results in increase of the thermal conductivity with the magnetic field. The effect stems from the gravitational anomaly appearing when chiral electrons propagate through curved space-time [17,19,22,71,72]. This violates separate conservation of energy-momentum tensor in a chiral system [21] and such a violation can be translated to thermal transport coefficients [19,20,22,72]. Astonishingly, the chiral heat current is related to the chiral electrical current by the Wiedemann-Franz law in the same way as the thermal and electric conductivities of free electrons [22,36,57]:

$$\kappa_{ch} = \frac{\pi^2 k_B^2 \sigma_{ch} T}{3e^2} = L_0 \sigma_{ch} T \qquad (2)$$

Where $L_0 = \pi^2 k_B^2/3e^2$, $k_B$ is the Boltzmann constant and $e$ is the elementary charge. In other words, a Weyl semimetal is not only expected to exhibit in-field increase of the thermal and electrical conductivities due to presence of the chiral anomaly, but these two transport coefficients should also obey the Wiedemann – Franz law. This is what we report to happen at low temperatures and high magnetic field in NdAlSi.

In summary, we have investigated the magneto-electrical and magneto-thermal transport in the antiferromagnetic Weyl semimetal NdAlSi. At low temperatures, we observed both positive magneto-electric and magneto-thermal conductivities, which appear to be related in the manner predicted by the Wiedemann – Franz law. This is the behaviour expected when additional electric and thermal currents contributions result from the quantum anomalies. The detected presence of the gravitational (or thermal chiral) anomaly in NdAlSi is a good confirmation of unusual quantum - based properties of Weyl semimetals.

**Methods**

Single crystals of NdAlSi were grown by a self-flux technique, details of which were described in the previous reports [39]. NdAlSi crystallizes in a non-centrosymmetric centred tetragonal structure, I4$_1$md ($C_{4v}$), but in case of site mixing between Al and Si, the space group could change from non-centrosymmetric to centrosymmetric. A neutron diffraction study was used to show that such a site mixing does not occur in NdAlSi [39].

For the transport measurements, a rectangular bar with dimensions 1.9 x 1.4 x 0.35 mm$^3$ was cut from a suitable single crystal with the longest side of the sample oriented along the [1 0 0] direction (crystallographic *a*-axis that is a magnetic hard axis) and the shortest side along the [0 0 1] direction (crystallographic *c*-axis that is an easy axis). The electrical resistivity (*ρ*) was measured using a four-point technique with an alternating electric current flowing along *a*. The current contacts were made on the cross-section surface rather than point-like to maintain the homogeneous current and minimize extrinsic effects. The experiments were performed in the temperature (*T*) range 1.8 - 300 K and in the magnetic field (*B*) up to 14.5 T applied parallel to the electric current (*B* ∥ *a*).

The isolated heater method was used for the thermal conductivity ($\kappa$) measurements, it was described in details in [16]. During measurements the thermal gradient ($\nabla T$) was applied along *a*-axis of the single crystal NdAlSi, whereas the magnetic field was parallel to the thermal gradient. For field sweeps, a DC technique was used and up and down sweeps of the magnetic field were performed to extract the field-symmetric component of the signal. For temperature ramps at constant field the quasi-AC mode was used.

**Acknowledgments**

This work was supported by the Foundation for Polish Science through the IRA Programme co-financed by EU within SG OP. The material is based upon work supported by the Air Force Office of Scientific Research under award number FA2386-21-1-4059. We thank Andrzej Szewczyk, Francisco Pena Benitez and Jakub Polaczyński for insightful discussions.

**Competing financial interests:**

The authors declare no competing financial interests.

**Data availability**

All of the relevant data that support the findings of this study are available from the corresponding author upon reasonable request.

**M.Matusiak ORCID iD**: 0000-0003-4480-9373

along the a-axis; fitting of magnetic field dependence of non-electronic thermal conductivity; evolution of the quantum oscillations with temperature; and data on the transverse magnetoresistance.

**Figures**

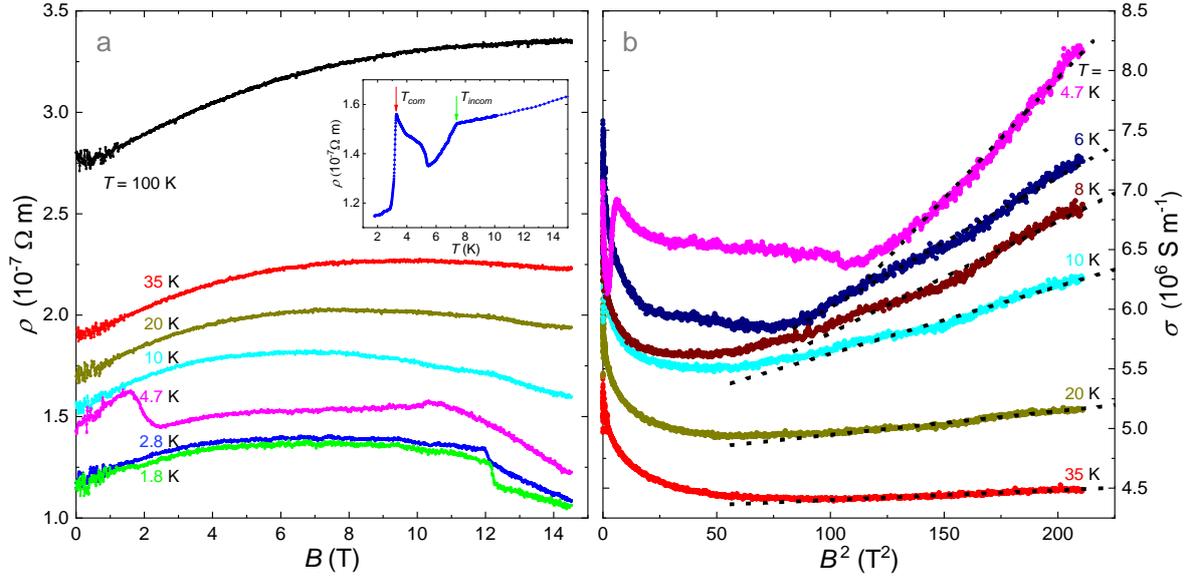

**Figure 1. Magnetic field dependences of the electrical resistivity and conductivity of NdAlSi measured with the electric current and magnetic field parallel to each other and oriented along *a*-axis.** Panel (a) presents $\rho(B)$ plots for several different temperatures. Inset in this panel shows the zero-field temperature dependence of the resistivity measured along [100] with anomalies at $T_{incom}$ = 7.2 K (green arrow) and $T_{com}$ = 3.2 K (red arrow) marking the incommensurate antiferromagnetic to commensurate ferromagnetic and paramagnetic to incommensurate antiferromagnetic transitions, respectively. Panel (b) presents Positive magneto-thermal conductivity plotted versus square of the magnetic field, where the dashed lines shows $B^2$ behaviour in the high field limit.

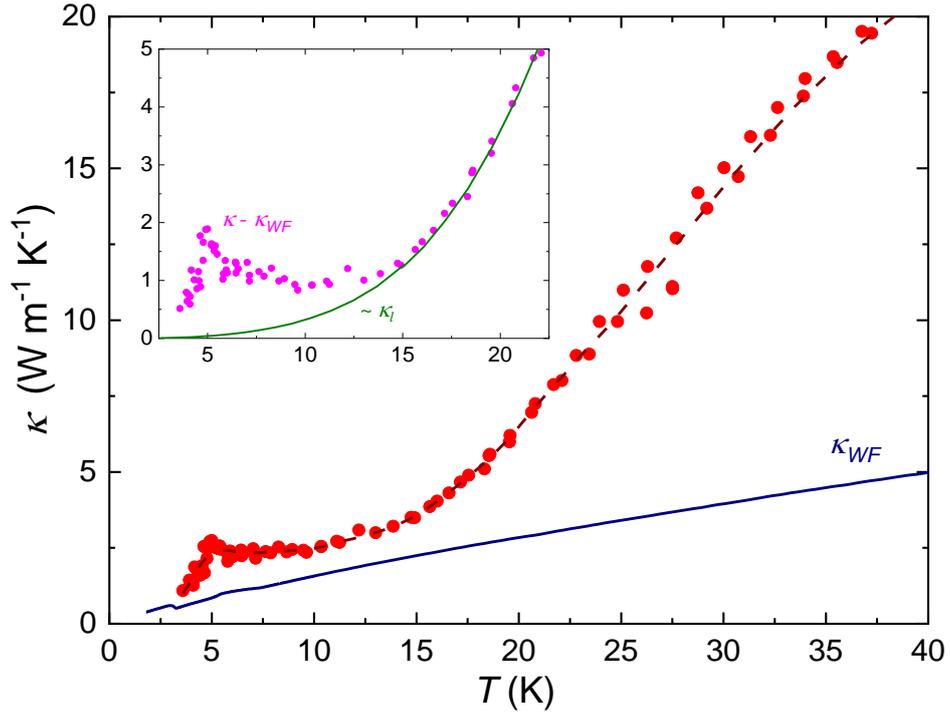

**Figure 2. The temperature dependence of the *a*-axis thermal conductivity for NdAlSi in zero magnetic field.** $\kappa(T)$ data is plotted with red points along with the electronic contribution $\kappa_{WF}(T)$ (dark blue line) calculated using the Wiedemann-Franz law. The dark red dashed line is a guide for the eye. Inset shows temperature dependence of the difference ($\kappa - \kappa_{WF}$) and phonon contribution to thermal conductivity $\kappa_l(T)$ (solid-green line) estimated using the LaAlSi specific heat data.

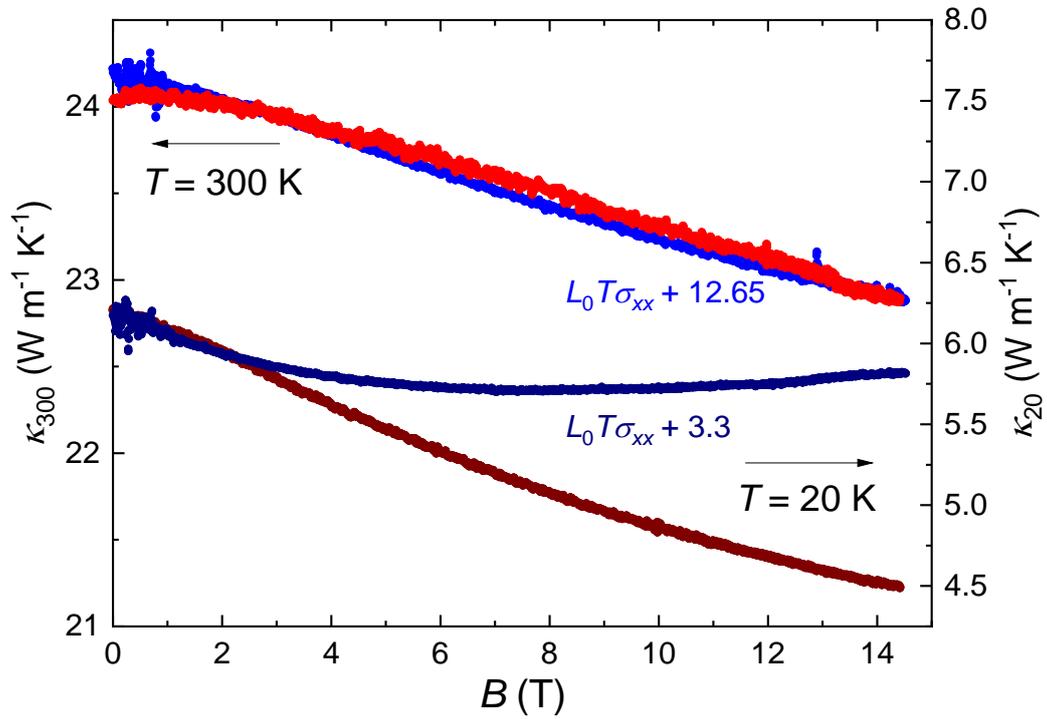

**Figure 3. The Wiedemann – Franz law.** The magnetic field dependences of the thermal conductivity $\kappa(B)$ for NdAlSi at $T = 20$ and $300$ K (plotted in dark red and red, respectively). These are compared with the corresponding $\kappa_{WF}(B)$ to which a presumably field independent phonon contribution has been added.

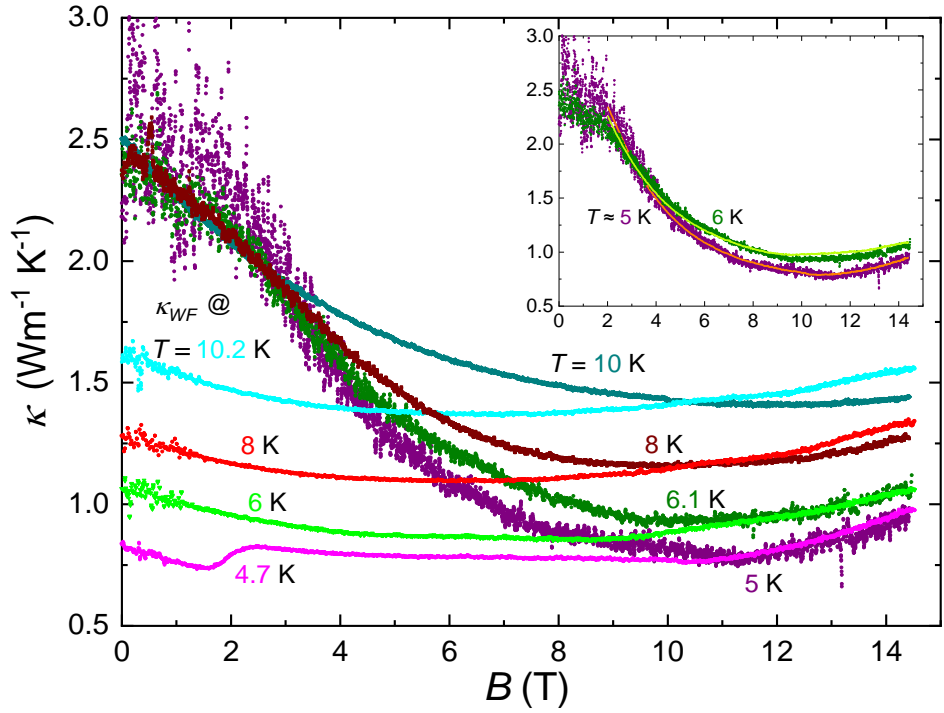

**Figure 4. Increase in the thermal conductivity of NdAlSi in the low temperature and high field range.** Dependences of the thermal conductivity on magnetic field compared to the corresponding $\kappa_{WF}(B)$. Inset shows $\kappa(B)$ at $T \approx 5$ and 6 K (purple and dark green, respectively) plotted with $\kappa_{WF}(B) + \kappa_m(B)$ (pink and green for $T \approx 5$ and 6 K, respectively), where $\kappa_m(B)$ is assumed to decay exponentially in the magnetic field (see Fig. S2 in SI).

# Supplementary Information

**Gravitational anomaly in ferrimagnetic topological Weyl semimetal NdAlSi**

Pardeep Kumar Tanwar [1], Mujeeb Ahmad [1], Md Shahin Alam [1], Xiaohan Yao [2], Fazel Tafti [2]

and Marcin Matusiak [1,3]

1. *International Research Centre MagTop, Institute of Physics, Polish Academy of Sciences, Aleja Lotnikow 32/46, PL-02668 Warsaw, Poland*
2. *Department of Physics, Boston College, Chestnut Hill, Massachusetts, 02467, USA*
3. *Institute of Low Temperature and Structure Research, Polish Academy of Sciences, ul. Okólna 2, 50-422 Wrocław, Poland*

**Temperature dependences of the thermal conductivity:**

The thermal conductivity ($\kappa$) of NdAlSi was measured in zero-field and for $B$ = 5, 10, 14.5 T applied along the *a*-axis. $\kappa(T, B=0T)$ exhibits a broad peak at $T \approx 60$ K, which appears to mark the temperature above which the Umklapp processes effectively disturb the thermal transport of phonons [1]. At lower temperature, $\kappa(T, B=0T)$ shows a hump at $T \approx 5$ K reflecting additional thermal transport by magnetic excitations. This hump is suppressed by the magnetic field and vanishes at $B = 10$ T.

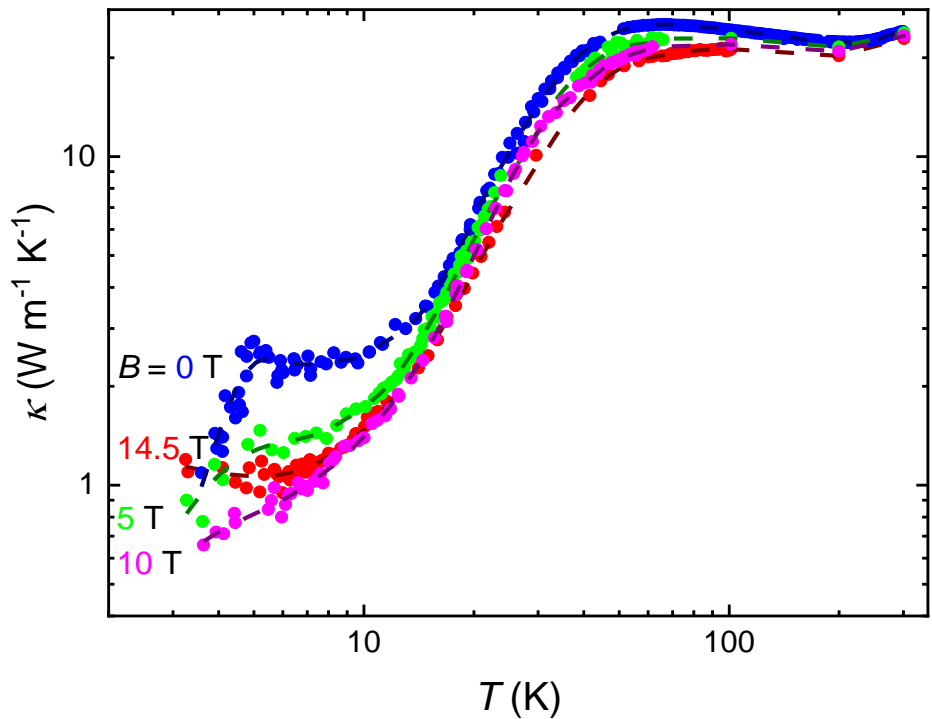

**Figure S1.** Temperature dependences of thermal conductivity of NdAlSi measured in the magnetic field ($B = 0$, 5, 10 and 14.5 T) parallel to temperature gradient applied along the crystallographic *a*-axis.

**Magnetic field dependence of the non-electronic thermal conductivity:**

At low temperature the non-electronic thermal conductivity quickly decreases with increasing $B$. Figure S2 presents the magnetic field dependence of thermal conductivity with subtracted electronic component ($\kappa_{WF}$) calculated using the Wiedemann – Franz law. The field dependence of $\kappa - \kappa_{WF}(B)$ can be well fitted with the exponential decay function $e^{-cB}$, which in the noninteracting approach is the expected behaviour of the magnonic thermal conductivity suppressed by a magnetic field [2,3].

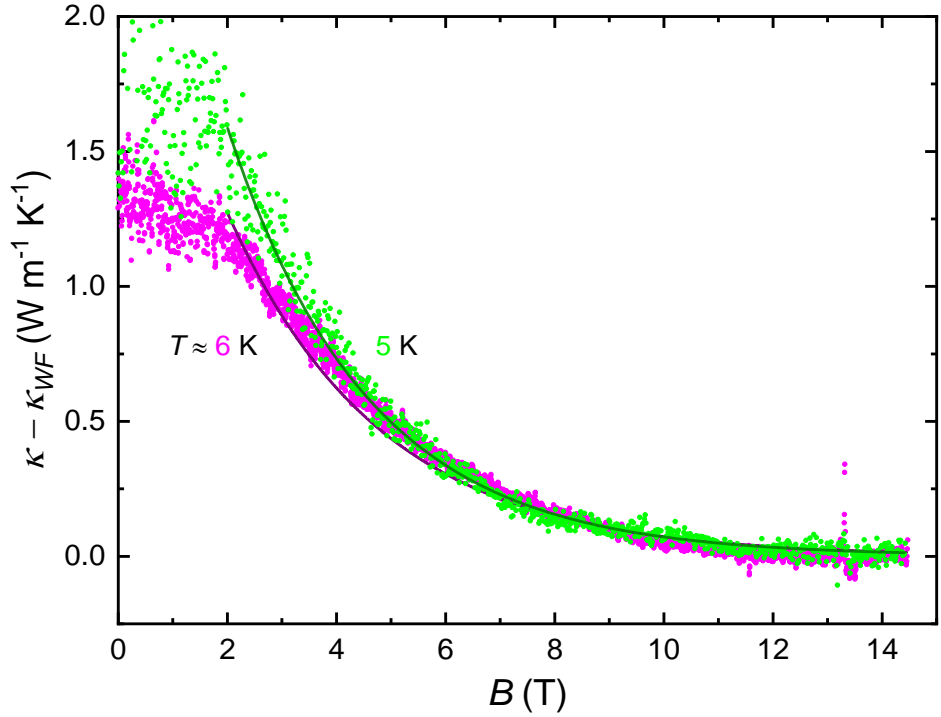

**Figure S2.** Magnetic field dependence of thermal conductivity of NdAlSi at $T \approx 5$ K (green points) and 6 K (magenta points) measured in $B \parallel \nabla T \parallel a$. The solid lines show the exponential decay function $be^{-cB}$; where $b = 3.45$ and 2.6 W m$^{-1}$ K$^{-1}$, while $c = 0.38$ and 0.35 for $T \approx 5$ and 6 K, respectively.

**Quantum Oscillations:**

The oscillatory component of the resistance ($\Delta R$) was extracted from $R(B)$ by subtracting the non-oscillatory background (in the form of a second order polynomial) in the magnetic field range 8 - 14.5 T. A distance between minima in $\Delta R(1/B)$ allows an estimate of the oscillation frequency: $F = 53$ T. The temperature dependence of the amplitude of this oscillations (disappearing at $T = 35$ K) was fitted with the thermal damping factor from the Lifshitz – Kosevich formula: $R_T = \frac{\alpha p X}{\sinh(\alpha p X)}$, where $\alpha = \frac{2\pi^2 k_B}{e\hbar}$, $p$ is the harmonic number, $X = \frac{m^* T}{B}$, $k_B$ is the Boltzmann constant, $e$: elementary charge, $\hbar$: reduced Planck constant. The resulting effective mass equals: $m^* = 0.11\ m_0$ ($m_0$ stands for the free electron mass).

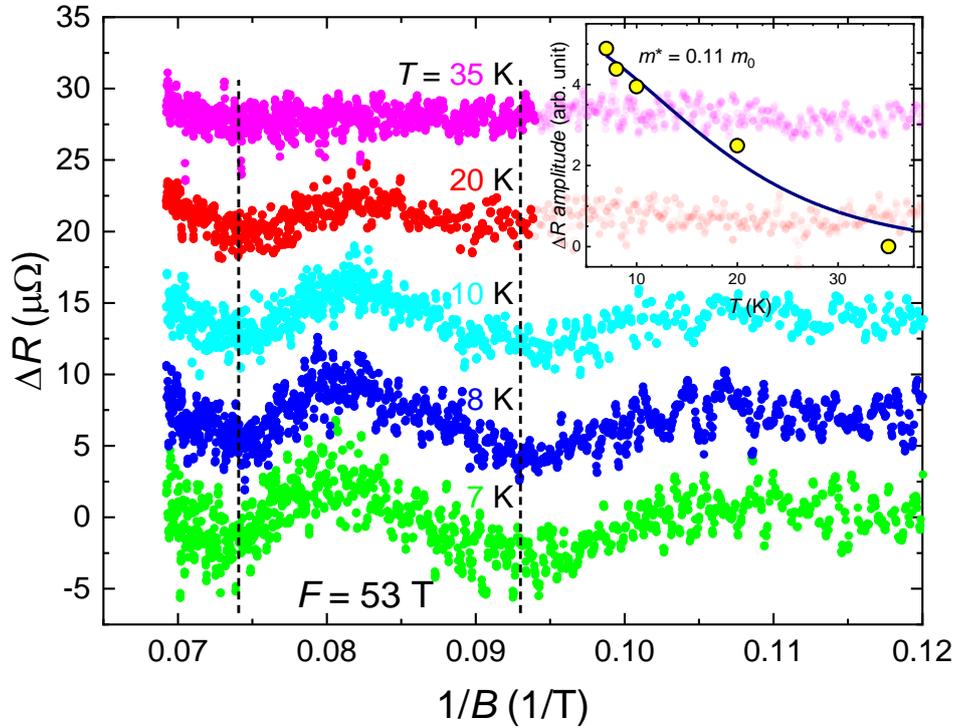

**Figure S3.** The magnetic field dependence of the oscillatory component of the NdAlSi resistance plotted for several temperatures. For temperatures $T = 8$ K and higher, each $\Delta R(1/B)$ dependence is shifted vertically (by 7 µΩ from previous one) for the sake of clarity. Inset shows the temperature dependence of the amplitude of the oscillations fitted with the function describing the thermal damping factor from the Lifshitz – Kosevich formula.

**Positive transverse magnetoresistance:**

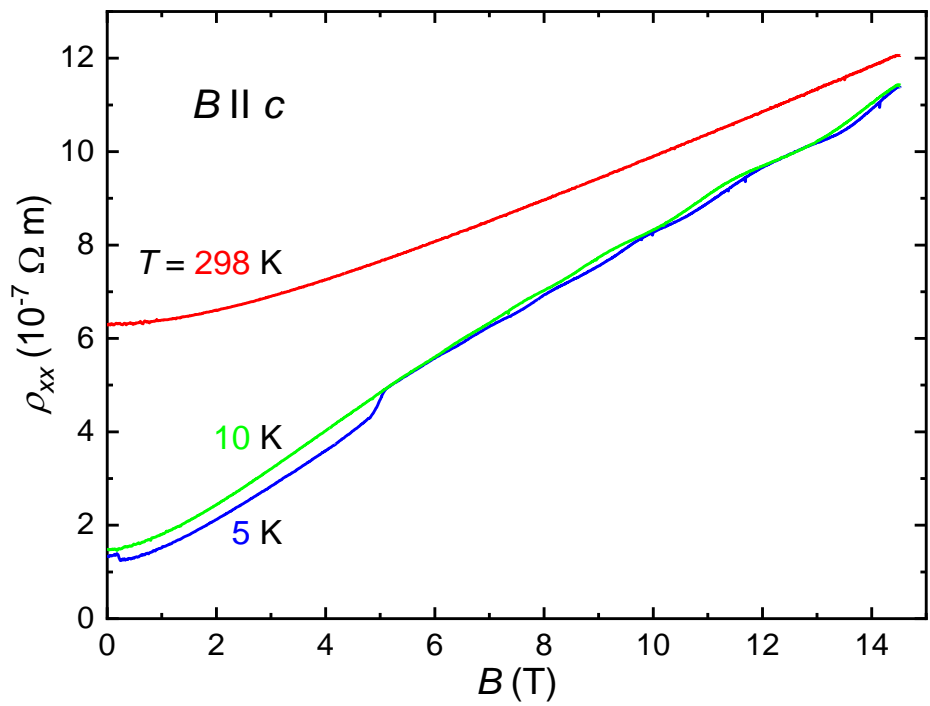

**Figure S4.** Magnetic field dependences of the electrical resistivity of NdAlSi measured at $T$ = 5, 10 and 298 K with the magnetic field oriented parallel to $c$-axis and perpendicular to the electric current.